\documentclass[12pt]{article}
\usepackage[T2A]{fontenc}
\usepackage{graphicx}
\usepackage[english]{babel}

\newcommand{\no}{\nonumber}
\newcommand{\non}{\nonumber \\}
\newcommand{\ve}[1]{{\bf #1}}
\newcommand{\be}{\begin{equation}}
\newcommand{\ee}{\end{equation}}
\newcommand{\bea}{\begin{eqnarray}}
\newcommand{\eea}{\end{eqnarray}}

\newcommand{\lb}{\left \{}

\newcommand{\rd}{\right .}

\newcommand{\vk}{\ve{k}}
\newcommand{\vl}{\ve{l}}
\newcommand{\vj}{\ve{j}}

\newcommand{\cM}{{\cal M}}

\newcommand{\cB}{{\cal B}}

\begin{document}
\title{Equation of State and  Thermodynamic Functions of the Ising-like Magnet at $T>T_c$}
\author{M.P. Kozlovskii,  O.O. Prytula}
\maketitle
\begin{center}
{\it Institute for Condensed Matter Physics of the  National Academy
of Sciences\\ of Ukraine,  1 Svientsitskii Str., Lviv 79011, Ukraine,\\ email: prytula@icmp.lviv.ua}
\end{center}

\abstract{
The $3D$ Ising-like  system in the external field is described using the non-perturbative collective variables method.
The universal as well as nonuniversal system characteristics are obtained within the framework of this approach.
The calculations are carried out on the
microscopic level starting from the Hamiltonian. They are valid in the whole $h-T$ plane of the critical region.
It is established, that the contributions related with wave vector values $\vk\rightarrow0$ exhibit the properties of the
total system near the critical point. The behaviour of the susceptibility as function of the temperature in
the presence of the field is investigated.
The locations of the maximums susceptibility on the temperature scale for different values of the field are established.
\\


\section{Introduction}
Despite the simplicity and clearness of the $3D$ Ising model, it is still the object investigated with help of
the approximate methods \cite{palvi102}. We propose the description of the behaviour of the
Ising-like magnet in the external field by collective variables (CV) method \cite{prfref5}. This approach is
si\-mi\-lar to the well-known non-perturbative Wilson-Pol\-chins\-kii Renormalization Group method \cite{btw02,tw94}.
In particular, the  coarse-grained procedure with using renormalization group (RG) transformation is
employed in the both methods. But, in contrast to the Wilson approach \cite{wk74}, the CV method does not involve any
phenomenological assumptions and adjusting parameters. It is based on the physically well-grounded and
mathematically rigorous use of the space of CV, which are associated with modes of spin moment density oscillations.
As result, the method allows one to obtain the macroscopic characteristics as functions of the microscopic parameters,
contained in the system Hamiltonian.

In the $3D$ spin systems exhibiting the critical behaviour, the presence of the asymmetry (presence of an external field)
sufficiently  complicates the problem. Usually, the series in scaling variable, which are some ratio of the field and
temperature variable, are used for description of the system behaviour. In this case, one of the factors dominates and
forms the system critical behaviour. Using CV method, we have already obtained such characteristics in the form
of a power series in the scaling variable in the regions of the weak and strong fields for
temperatures above and below $T_c$ ($T_c$ is the phase transition temperature in the absence of an
external field) \cite{kppfer05,kpprevb06}.
When the role of the field and temperature is equally important, the series are
not valid. For complete system description,  it is necessary to perform calculations in the whole $h-T$ plane,
where $T$ is the temperature, and $h$ is an external field.

In the perturbative approaches ($\epsilon$-expansion, high-tem\-pe\-ra\-tu\-re expansion, etc), the parametric
equation of state is used for description such systems \cite{gz197}. Within the framework of the Wilson-Polchynskii method,
the investigation of the $3D$ Ising model critical behavior at the presence of an external field is performed in
the so-called local potential approximation (LPA) \cite{t02}. In particular, the critical exponents of the correlation
length $\nu$ and order parameter $\delta$ as well as asymmetric leading corrections to scaling related with presence of the field
are calculated. The description of such a system can be used also for investigations of the nonmagnetic systems. For instance,
the critical behaviours of the system "gas-liquid" in the  plane (temperature,pressure) \cite{palvi102}, Standard Model of
the electroweak interactions in the plane (higgs mass, temperature) \cite{rtkls98,klrs96}, were shown to have similar features
of the phase transition. These models belong to the $3D$ Ising universality class.

\section{Basic relations}
We represent the calculations, which are valid in the whole $h-T$ plane of the critical region.
Starting with the Hamiltonian
\be
H=-\frac{1}{2}\sum_{\vl,\vj}\Phi(r_{\vl \vj})
\sigma_\vl\sigma_{\vj}-h\sum_{\vl}\sigma_{\vl},
\label{f1}
\ee
and using the "layer-by-layer" integration of the partition function, we obtain the critical exponents, free energy and other
thermodynamic and structural characteristics of the system. The quantities $\sigma_{\vl}$ are
$z$-components of the spin operator, whose eigenvalues take on the values $\pm1$, and $h$ is an external field.

The interaction between spins on the simple cubic lattice is described by the exponentially
decreasing function  $\Phi(r_{\vl \vj})=A\exp(-r_{\vl\vj}/b)$ of the distance $r_{\vl\vj}$ between particles
at sites $\vj$ and $\vl$. Such an approach makes it possible to investigate the system behaviour
depending on the microscopic parameters, in particular, on the radius of the effective
interaction $b$ and constant $A$.

For the following calculations, it is necessary to perform the transition from quantities $\sigma_{\vl}$
to the CV $\rho_{\vk}$ through the functional representation for operators of the spin density oscillation modes
\[
\hat{\rho_\vk}=(\sqrt{N})^{-1}\sum_{\vl}{\sigma_{\vl}\exp(-i\vk\vl)}.
\]
Here $N$ is the number of particles in the system. The detailed procedure of this transition is described in
works \cite{novo89,prfref5}.
We use also the parabolic approximation of the Fourier transform of the interaction potential in the form
\be
\Phi(k)=\lb
\begin{array}{ll}
\Phi(0)(1-2b^2k^2), & k\leq B_0,\\
\Phi_0=\Phi(0)\bar\Phi, & B_0<k\leq B,
\end{array}
\rd \label{f2}
\ee
where $B=\pi/c$ is the boundary of the  Brillouin half-zone, and the quantity $B_0=B/s_0$
and $s_0\geq2$ is the starting division parameter. This parameter defines the boundary of the region,
where the parabolic approximation of the interaction potential is valid.
The quantity $\bar\Phi$ is the constant and $\Phi(0)=8\pi A(b/c)^3$, where $c$ is the lattice
constant \cite{k05}. We begin the calculation with partition function \cite{prfref7,k05}
\bea
Z&\propto&\int (d\rho)^N_0(d\omega)^N_0 \exp \Biggl[ \frac{1}{2}\beta
\sum_{\vk\in\cB_0}(\Phi(k)-\Phi_0)\rho_{\vk}\rho_{-\vk}\Biggr]\non
&&\times\exp(2\pi i\sum_{\vk\in\cB_0}{\rho_\vk}\omega_{\vk})Z(\omega).
\label{f3}
\eea
Here  $\beta=1/kT$ and $k$ is the Boltzmann constant.
The quantity
\bea
Z(\omega)&=&\exp \Biggl\{\sum_{n\geq0}[\frac{(-2\pi i)^n}{n!}N^{1-n/2}\non
&&\times\cM_n(h')
\sum_{\vk\in\cB}{\omega_{\vk_1}}...\omega_{\vk_n}\delta_{\vk_1+...+\vk_n}]\Biggr\}
\label{f4}
\eea
is the
the cumulant expansion of the Jacobian of transition from the set of the spin variables
$\sigma_{\vl}$ to the set of CV $\rho_{\vk}$, where $\delta_{\vk_1+...+\vk_n}$ is the Kronecker symbol.
The coefficients $\cM_n$ are defined with help of the relations:
\be
\cM_0=\ln\cosh{h'}, \;\;\; \cM_n=\frac{\partial^n\cM_0}{\partial h'^n}.
\label{f5}
\ee
The integer even values of the quantity $n$ in formulas (\ref{f4}) and (\ref{f5}) determine the choice of the effective potential (models $\rho^4$, $\rho^6$, etc),
which will be used for the description of the system.
The value $n=0$ corresponds to the mean-field approximation, and at $n=2$, we have Gauss distribution of the fluctuations
in the system. The odd degrees of CV  appear as result of the presence of an external field.
In this work, the calculations are carried out employing the simplest non-Gaussian model with $n=4$.
The results allow one to take into account long-range interactions, which determine the critical behaviour of the system.
It should be noted, that the variable $\rho_0$, which corresponds to the zero value of the wave vector, is crucial
in the forming of the partition function (\ref{f3}). The average value of this quantity is the order parameter of the
system \cite{prfref5}.

For the integration of (\ref{f3}), one should pass to the  following field variables by performing
the substitution:
\bea\rho_{\vl}=(\sqrt{N_0})^{-1}\sum_{\vk\in\cB_0}{\rho_{\vk}\exp(i\vk\vl)}, \non
\omega_{\vl}=(\sqrt{N_0})^{-1}\sum_{\vk\in\cB_0}{\omega_{\vk}\exp(-i\vk\vl)} \no,
\eea
where $N_0=Ns_0^{-3}$. In this case, the partition function takes on the
form \cite{k05}:
\bea
Z&=&Z_0 j_0 \int (d\rho)^N_0(d\omega)^N_0 \exp \Biggl[ \frac{1}{2}\beta
\sum_{\vk\in\cB_0}(\Phi(k)-\Phi_0)\rho_{\vk}\rho_{-\vk}\Biggr]\non
&&\times\prod_{\vl}{I_{\vl}(\rho_{\vl})}.
\label{f6}
\eea
The vector $\vl$ varies in the volume of periodicity $V=N_0c_0^3$, where $c_0=s_0c$, and
\[
Z_0=2^N \exp{[N(\frac{1}{2}\beta\Phi_0+\cM_0)]}, \qquad j_0=\sqrt2^{N_0-1}.
\] It should be noted, that the
variable $\omega_{\vl}$ can be interpreted as the scalar field, which is similar to the Habbard-Stratanovich
field. Within the frames of the quartic model, the integrals $I_{\vl}(\rho_{\vl})$ is defined by the expression
\bea
I_{\vl}(\rho_{\vl})&=&\frac{1}{\pi\sqrt2}e^{-a_h^{''}}\int d\omega_{\vl}\exp[2\pi i\rho_{\vl}(\frac{\omega_{\vl}}{\pi\sqrt2}-
i\frac{s_0^{d/2}h'}{2\pi})]\non
&&\times\exp[\omega_{\vl}^2-c_h^{''}\omega_{\vl}^4].
\label{f9}
\eea
Here $a_h^{''}=s_0^dh'^2/2$ and $c_h^{''}=s_0^{-d}(1-4h'^2)$.
We perform also the substitution of variables to cancel the cubic term in the cumulant expansion.
The coefficients in the exponent expression are approximated by  power  series in field variable
to $h'^2$. This approximation is quit precise, since in the vicinity of the critical point,
the maximum values of the field are of order of $0.01$. Such values of the field correspond to the
real fields, which are commensurate with fields of the order parameter saturation of the ferromagnetic materials.
We will find the result of the integration of the expression (\ref{f9}) in the form
\be
K_{\vl}(\rho_{\vl})=e^{a_0}\exp\Biggl[-\sum_{n=1}^4\frac{a_n}{n!}\rho_{\vl}\Biggr],
\label{f10}
\ee
where the coefficients  $a_n$ are defined using the condition
\[
\frac{\partial^nI_{\vl}}{\partial \rho_{\vl}^n}\Bigg|_{\rho_{\vl}=0}=\frac{\partial^nK_{\vl}}{\partial \rho_{\vl}^n}\Bigg|_{\rho_{\vl}=0}.
\]
Since the quantity  $c_h^{''}\ll1$, we present the integrand of the (\ref{f9}) by the power series in the quartic
term. Such an approximation  is used only in the space of variables $\rho_{\vl}$. For the coefficients $a_n$, we have
\bea
a_0=\ln(\sqrt\pi(1-\frac{3}{4}c_h^{''})), \non
a_1=-s_0^{d/2}h',\;\;\;
a_2=1-3c_h^{''},\non
a_3=0,\;\;\;
a_4=6c_h^{''}.
\label{f11}
\eea
In the space of the variables $\rho_{\vk}$, the partition function takes on the form:
\bea
Z_0&=&Z_0j_0(\pi\sqrt2)^{-1}e^{(-a_h^{''}+a_0)N_0}\int(d\rho)^{N_0}\exp[-a_1\sqrt N_0\rho_0\non
&&-\frac{1}{2}\sum_{\vk\in\cB_0}d(k)\rho_k\rho_{-k}\non
&&-\frac{a_4}{4!N_0}\sum_{\vk\in\cB_0}\rho_{k_1}...\rho_{k_4}\delta_{k_1+...+k_4}],
\label{f12}
\eea
where $d(k)=a_2+\beta\Phi_0-\beta\Phi(k)$.
The presence of the external field is represented only by the linear term in the exponent of the expression (\ref{f12}).
It should be stressed that the same result is valid for the models with $n=6,8,...$.
The field dependence of other coefficients is very weak in comparison to coefficient in the linear term.

The partition function (\ref{f12}) is starting point to perform the "layer-by-layer" integration with using
RG transformations for the quantities $a_n$. Beginning the integration from the variables $\rho_{\vk}$ with
large values of ${\vk}$, we divide the the phase space of the CV $\rho_{\vk}$ into the layers with the
division parameter $s$. In each $n$th layer (where $B_{n+1}<k<B_{n}$, $B_{n+1}=B_n/s$) the Fourier
transform of the potential $\Phi(k)$ is replaced by its average value. Such an approximation corresponds
to the LPA in the Wilson-Polchinskii nonperturbative theory \cite{btw02}.
The detail description of this procedure is given in the paper
\cite{k05}. The result of the integration can be represented by the expression
\be
Z=Z_0Q_0Q_1...Q_n[Q(P^{(n)})]^{N_{n+1}}I_{n+1},
\label{f13}
\ee
where $Q_0=[(\pi\sqrt2)^{-1}e^{-a_h^{''}+a_0}Q(d)]^{N_0}$, and quantities
\be
Q_n=[Q(P^{(n-1)})Q(d_n)]^{N_n}
\label{f14}
\ee
are the partial partition functions of the   $n$th block lattice. Here
\bea
Q(P^{(n-1)})=(2\pi P_2^{(n-1)})^{-\frac{1}{2}}\Bigl(1-\frac{3}{4}G^{(n-1)}\Bigl),\non
Q(d_n)=\Biggl(\frac{24}{a_4^{(n)}}\Biggr)^{\frac{1}{4}}\gamma_1\Bigl(1-\gamma h_2^{(n)}\Bigl)
\label{f15}
\eea
as well as
\bea
P_2^{(n-1)}=\Biggl(\frac{24}{a_4^{(n-1)}}\Biggr)^{\frac{1}{2}}\gamma\Bigl(1+t_2 h_2^{(n-1)}\Bigl),\non
G^{(n-1)}=s^{-d}G_0(1+G_2h_2^{(n-1)}),\non
h_2^{(n)}=\sqrt6 \frac{d_n(B_n,B_{n+1})}{(a_4^{(n)})^{{1/2}}}.
\label{f16}
\eea
The quantities $\gamma$, $\gamma_1$, $G_0$, $t_2$ and  $G_2$ do not depend on the field. They are
given in \cite{k05}. The coefficient
\[d_n(B_n,B_{n+1})=d_n(0)+qs^{-2n}
\]
is defined through the
value, which related to the interaction potential $\beta\Phi(B_n,B_{n+1})$ averaged on the
interval $\vk\in\cB_n\backslash \cB_{n+1}$. As result, we have
\be
d_n(k)=a_2^{(n)}+\beta\Phi(B_n,B_{n+1})-\beta\Phi(k).
\label{f17}
\ee
Here $q=\bar q\beta\Phi(0)$ and $\bar q=(b\pi/c)^2s_0^{-2}(1+s^{-2})$.
The integral  $I_{n+1}$ in the formula (\ref{f13}) has the
following form:
\bea
I_{n+1}&=&\int (d\rho)^N_{n+1}\exp\Biggl\{-a_1^{(n+1)}\sqrt N_{n+1}\rho_0\non
&& -\frac{1}{2}\sum_{\vk\in\cB_{n+1}}d_{n+1}(k)\rho_{\vk}\rho_{-\vk}\non
&&-\frac{a_4^{(n+1)}}{4!N_{n+1}}\sum_{\vk\in\cB_{n+1}}\rho_{\vk_1}...\rho_{-\vk_4}\delta_{\vk_1+...+\vk_4}\Biggr\}.
\label{f18}
\eea
The coefficients $a_l^{(n+1)}$ are defined by the recurrent relations (RR)
\bea
&&a_1^{(n+1)}=s^{d/2}a_1^{(n)}\non
&&a_2^{(n+1)}=f_{00}(a_4^{(n)})^{1/2}(1+\alpha_2h_2^{(n)})\non
&&a_4^{(n+1)}=s^{-d}f_{01}a_4^{(n)}(1+\alpha_4h_2^{(n)}),
\label{f19}
\eea
obtained as result of the RG transformations. The quantities $\alpha_2$, $\alpha_4$, $f_{00}$ and $f_{01}$
are given in \cite{k05}. Due to the presence of the field, we obtain $l+1$ RR for the models
$\rho^{2l}$, where $l=1,2,3..$.
Performing the transformation $\rho_{\vk}=s \rho_{\vk}'$, introducing the notation $\omega_n=s^na_1^{(n)}$,
$r_n=s^{2n}d_{n}(0)$, $u_n=s^{4n}a_4^{(n)}$ and taking into account the expression for $h_2^{(n)}$ in (\ref{f16}),
the equation (\ref{f19}) can be written in the matrix form
\be
\\
\left( \begin{array}{c}
\omega_{n+1}-\omega^* \\
r_{n+1}-r^*\\
u_{n+1}-u^*
\end{array} \right) =
\left( \begin{array}{ccc}
R_{11} & 0 & 0 \\
0 & R_{22} & R_{24}\\
0 & R_{42} & R_{44}\end{array} \right)
\\
\left( \begin{array}{c}
\omega_n-\omega^* \\
r_n-r^*\\
u_n-u^*
\end{array} \right),
\label{f20}
\ee
where $R_{11}=s^{\frac{d+2}{2}}$, $R_{22}=s^2f_{00}\alpha_2\sqrt6$, $R_{24}=s^2f_{00}u^{*-1/2}$,
$R_{42}=sf_{01}\alpha_4\sqrt6u^{*1/2}$ and $R_{44}=sf_{01}$. The quantities $w^*=0$, $r^*=-q$ and $u^*=q(1-s^{-2})/f_{00}$
are the fixed point coordinates. For parameter $s$, there exists a preferred value $s=s^*$, which  is
determined by condition $h_2^{(n)}=h_2^{(n+1)}=0$.

\section{RR Solutions. Scaling Region.}
The fixed point is associated with the phase transition point, since in this case, the fluctuations become
infinitely large, and it is necessary to perform infinite number of iterations for system description.
But at the fixed values of the field $h'$ and reduced temperature $\tau=(T-T_c)/T_c$, this number become finite.
Thus, one can linearize the RR (\ref{f19}) in the case of the small deviations from the fixed point and find the
coefficients of the $n$th block structure through their initial values (\ref{f11}).
Hence, one can find the eigenvalues and eigenvectors of the RG matrix and write down the solutions of the equation
(\ref{f20}) in the following form:
\bea
& & \omega_n=\omega^*-s_0^{d/2}h'E_1^n,\non
& &r_n=r^*+c_{k1}^{(0)}\beta\Phi(0)\tau E_2^n+ c_{k2}T_{24}^{(0)}
(\varphi_0^{1/2}\beta\Phi(0))^{-1}E_4^n,\non
& & u_n=u^*+c_{k1}^{(0)}(\beta\Phi(0))^2T_{42}^{(0)}\varphi_0^{1/2}\tau E_2^n+c_{k2}E_4^n. \label{f21}
\eea
Here  $E_1=20.98$, $E_2=7.37$, $E_4=0.4$ are the eigenvalues of the RG matrix,
$T_{24}^{(0)}=u^{*1/2}R_{24}(E_4-R_{22})^{-1}$,
$T_{42}^{(0)}=u^{*-1/2}R_{42}(E_2-R_{44})^{-1}$. The coefficients $c_{k1}^{(0)}$ and $c_{k2}$
are expressed by relations
\bea
c_{k1}^{(0)}&=&\Bigl[1-\bar\Phi-\bar q-T_{24}^{(0)}u_0\varphi_0^{-1/2}(\beta\Phi(0)\beta_c\Phi(0))^{-1}\non
&&-T_{24}^{(0)}\varphi_0^{1/2}\Bigr]\times\Bigl[1-T_{24}^{(0)}T_{42}^{(0)}\Bigr]^{-1},\non
c_{k2}&=&\Bigl[u_0-u^*-T_{42}^{(0)}\varphi_0^{1/2}\beta\Phi(0)(r_0-r^*)\Bigr]\non
&&\times\Bigl[1-T_{24}^{(0)}T_{42}^{(0)}\Bigr]^{-1},\no
\eea
where $\varphi_0=(\bar q(1-s^{*-2})/f_{00})^2$. These values do not depend on the field.

Using the linearity condition for relations (\ref{f21}), we define the number $n$ of the
iterations. This is so-called  scaling region, which is defined by the values of the field $h'$
and temperature $\tau$.
The deviations from the fixed point are formed mainly in the first two equations of the system (\ref{f21}),
since in the last equation for $u_n$, the coefficient near the quantity $\tau E_2^n$ is small in comparison
to the coefficient near the same quantity in the equation for $r_{n}$. Taking into account above mentioned
condition, we write down the equation
\be
(-s_0^{d/2}h'E_1^{n_p+1})^2+(c_{k1}^{(0)}\tau \beta\Phi(0)E_2^{n_p+1})^2=r^{*2},
\label{f22}
\ee
which allows us to find the number $n_p$ of the "layer-by-layer" integrations.
The definition of the quantity $n_p=n_p(\tau,h')$ as function of the field and temperature is essential for
investigating the crossover between temperature-dependent and field-dependent critical behaviour
of the system. Indeed, this number determines the size of the block structure, which is commensurate
with correlation length. Equality of the terms in the left side of the equation (\ref{f22}) corresponds
to the state of the system in the vicinity of the line  determined by relation
\be
\tau\propto h^{\frac{1}{\beta\delta}},
\label{f23}
\ee
where $\beta$ and $\delta$ are the critical exponents of the order parameter. In this case, the field and temperature
equally influence on the forming of the critical behaviour of the system.

\section{Universal and Nonuni\-ver\-sal cha\-rac\-te\-ris\-tics of Sys\-tem}

In the high-temperature region, the partition function is represented as
\be
Z=Z_0Q_0Q_1...Q_{n_p}Q_{n_p+1}[Q(P^{(n+1)})]^{N_{n_p+1}}I_{n_p+2}.
\label{f24}
\ee
It should be noted, that in comparison to (\ref{f11}), the expression  (\ref{f24}) contains additional
partial partition function $Q_{n_p+1}$. Its presence is related with performing the additional step of
integration. Such an integration is carried out for reaching the dominant value of the coefficient $d_{n_p+2}$ in
comparison to other coefficients in $I_{n_p+2}$ for any $\vk\neq0$, where $\vk\in\cB_{n+2}$ (see (\ref{f18})).
Thus, for these values of the wave vector, we perform the integration with respect to the collective variables
in the Gauss approximation.

Using the partition function (\ref{f24}), we  represent the system free energy
as
\be
F=F_{\vk\neq0}+F_{\vk=0}.
\label{f25}
\ee
The second term in this expression is related to the integration of the quantity
\bea I_{n_p+2}^{(0)}&=&\int (d\rho_0)\exp \Biggl[N^{1/2}
h'\rho_0-\frac{1}{2}d_{n_p+2}(0)\rho^2_0\non
&&-\frac{1}{4!}a_4^{(n_p+2)} N^{-1}_{n_p+2}\rho^4_0\Biggr]
\label{f26}
\eea
with respect to the variable $\rho_0$ using the steepest-descent method.
As was mentioned above, this variable cor\-res\-ponds to the order parameter and forms the
main contribution to the free energy of the system. Using the substitution of variables
$\rho_0=\sqrt N\bar\rho_0$ and the maximum condition for the expression in the exponential function
(see (\ref{f26})) by differentiating with respect to the variable $\bar \rho_0$, we find the root
of the cubic equation in the form
\be
\bar\rho_0=\sigma_0s^{-\frac{n_p+1}{2}}.
\label{f27}
\ee
The choice of the valid root is based on the free energy minimum condition.
The term $F_{\vk=0}$ is defined by expression
\bea
F_{\vk=0}&=&-NkT s^{-3(n_p+1)} \Bigl[h'\sigma_0 E_1^{n_p+1}-\frac{1}{2}r_{n_p+2}s^{-2}\sigma^2_0 \non
&&-\frac{1}{4!}s_0^{d}s^{-1}u_{n_p+2}\sigma^4_0\Bigr].
\label{f28}
\eea
The expression (\ref{f27}) for $\bar \rho_0$ describes the order parameter behaviour
and can be interpreted as equation of state.

For another term in the right side of (\ref{f25}), we have the relation
\bea
F_{\vk\neq 0}&=&-NkT \Bigl[\ln\cosh h'+l_0+l_2\tilde h^2\non
&&+l_3\tilde\tau+l_4\tilde\tau^2+l_{1Te}^{{(0)}}s^{-3(n_p+1)}\Bigr],
\label{f29}
\eea
which is the result of the integration with respect to the variables related with
nonzero values of the wave vector. Here, the coefficients
 $l_0$ , $l_{1Te}^{(0)}$, $l_2$, $l_3$ and $l_4$ are independent of the field  \cite{kpppr04},
 $\tilde h=h'/\bar q$, $\tilde \tau=c_{k1}^{(0)}\tau/\bar q$.
The contribution (\ref{f29}) is small in comparison to  $F_{\vk=0}$.
Indeed, its magnitude depends on the accuracy of the integration.
With increasing the accuracy of the calculations, this contribution vanishes.
It is physically grounded, since as result of the RG transformations, we deals with larger
lattice structure, which exhibits the properties of the previous smaller ones. Thus, the result of the last
integration with respect to the variable $\rho_0$ should represent the properties of the whole system.
In the figure \ref{fff1}, the average value of the variable $\rho_0$ and order parameter obtained
by differentiating the free energy $F$ with respect to the field as functions of the field are represented.
As one can see, the difference between these quantities is neglible small.
The result for the order parameter agrees with the data obtained by the Monte-Carlo simulations \cite{ts194}.
Some deviation in the strong field region is explained by the insufficient accuracy of the model $\rho^4$.
As has been  shown in the similar investigations in the absence of the field, the better results can be obtained
with using the higher approximations for the effective potential, in particular, $\rho^6$ model
\cite{prfref7,ypk02,ypk022}.

Thus, the expression (\ref{f28}) represents the total free energy of the system obtained using the sequential
mathematical transformations starting from the Hamiltonian (\ref{f1}). The relation (\ref{f27}) is the equation
of state, which is valid in the whole h-T plane.

In the case of  $h=0$, taking into account the solution of (\ref{f22}) for $n_p$, the free energy takes on the form:
\be
F_{\vk=0}= -NkT R\tilde\tau^{3\nu},
\label{f30}
\ee
where $R$ is the sum of the last two terms in (\ref{f28}). The critical exponent of the correlation length
satisfies the relation
\be
\nu=\frac{\ln s^*}{\ln E_2}\approx 0.61.
\label{f31}
\ee
The equation of state (\ref{f27}) is transformed into expression
\be
\bar\rho_0=\sigma_0\tilde\tau^{\frac{\nu}{2}}.
\label{f32}
\ee
At $T=T_c$, the second term in the (\ref{f22}) disappears. In this case, the free energy can be written down as
\be
F_{\vk=0}= -NkT R' \tilde h'^{\frac{6}{\delta}},
\label{f33}
\ee
where $R'=\sigma_0r^*/s_0^{d/2}-r^*s^{-2}\sigma^2_0/2-s_0^{d}s^{-1}u^*\sigma^4_0/4!$ and $\tilde h'=h's_0^{d/2}/r^*$.
Correspondingly to (\ref{f27}), the equation of state  has the form
\be
\bar\rho_0=\sigma_0\tilde h'^{\frac{1}{\delta}}.
\label{f34}
\ee
The critical exponent of the order parameter takes on the value  $\delta=5$.
Within the frames of the LPA of Wilson-Polchinskii theory, the values of critical exponents of the correlation
length and order parameter are   $\nu=0.6895$ and $\delta=5$ respectively \cite{t02} (one of the best estimates
for the exponent of the correlation length is $\nu=0.63002$ \cite{cprv199}).
Using the expression (\ref{f28}) for the free energy, we calculate also the system susceptibility $\chi$.
In the figure \ref{fff2}, it is shown the dependence of this characteristic on the temperature for different values
of the field. The maximums of sussepibility are located at the value of the scaling variable
$\sigma_m=\tau_m/h^{1/\beta\delta}\approx0.77$.
The same quantity in the mean-field approximation takes on the value
$\approx 0.8255$ \cite{mfw196}.

\section{Conclusions}
The general description of the $3D$ Ising-like system in the external field by the CV method is represented.
The approach is nonperturbative and can be used for the various systems, for which the perturbative methods
are not quit appropriated. It allows one to calculate macroscopic characteristics of the system derived from
the microscopic quantities. The investigations are performed in the whole h-T plane of the critical region.
The main advantage of the CV approach in comparison to others is the absence of any phenomenological assumptions
and adjusting parameters. All calculations are carried  out on the microscopic level.

Using the coarse-graining procedure, we obtain the explicit analytic expressions for the  free energy and order
parameter of the system. It is established, that the contribution related to the $\rho_0$ variable
exhibits the properties of the total system. Another part corresponding to the nonzero values of the wave vector
is essentially small and depends on the accuracy of the integration.
The crossover to the limit cases  ($h=0$, $T\neq T_c$) and ($h\neq0$, $T=T_c$) is demonstrated. The critical exponents
of the order parameter and correlation length are calculated.
The dependence of the order parameter as function of the external field is plotted. The result agrees with the result, obtained
by Monte-Carlo simulations, in the weak-field and crossover region. Some deviations in the region of the strong fields
can be explained by the quartic approximation of the effective potential. In order to obtain more accurate results,
it is necessary to employ the approximations with $n=6,8,..$.
The system susceptibility is calculated and plotted as function of the temperature for different values of the field.
We investigate the maximum location of this function. It is shifted to the region of the stronger fields
in comparison to result  obtained using the mean-field approximation \cite{mfw196}.

The investigations can be extended to the model with the n-component order parameter as it was shown in \cite{yuk100}
in the case of the absence of the field. The proposed approach permits one to study also the systems with
limit number of the particles.

% BibTeX users please use
% \bibliographystyle{}
% \bibliography{}
%
% Non-BibTeX users please use

\begin{figure}[htb]
%\resizebox{0.5\columnwidth}{!}
\centerline{\includegraphics[width=0.6\textwidth]{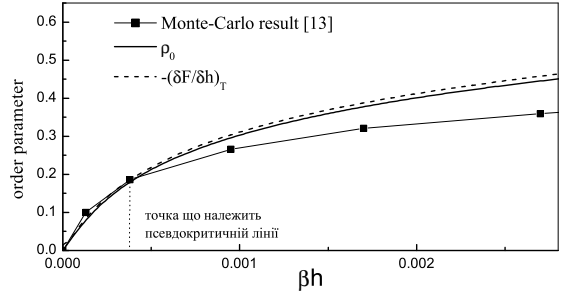}}
\caption{The order parameter of the system as function of the external field at $\tau=5\cdot10^{-3}$.} \label{fff1}
\end{figure}

\begin{figure}[htb]
%\resizebox{0.5\columnwidth}{!}
\centerline{\includegraphics[width=0.6\textwidth]{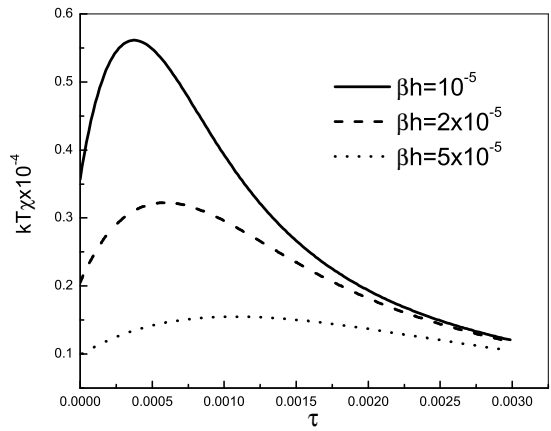}}
\caption{The system susceptibility as function of the temperature for different values of the field.}
\label{fff2}
\end{figure}

\end{document}